\documentclass[12pt,preprint]{aastex}
\usepackage{graphicx}

\begin{document}

\title{The Highest Resolution {\em Chandra} View of Photoionization and Jet-Cloud Interaction in the Nuclear Region of NGC 4151}

\author{Junfeng Wang, G. Fabbiano, M. Karovska, M. Elvis, G. Risaliti and A. Zezas}
\affil{Harvard-Smithsonian Center for Astrophysics, 60 Garden St, Cambridge, MA 02138} 
\email{juwang@cfa.harvard.edu; gfabbiano@cfa.harvard.edu; mkarovska@cfa.harvard.edu; elvis@cfa.harvard.edu; risaliti@cfa.harvard.edu; azezas@cfa.harvard.edu}

\author{C. G. Mundell}
\affil{Astrophysics Research Institute, Liverpool John Moores University, Birkenhead CH41 1LD, UK}
\email{cgm@astro.livjm.ac.uk}

\begin{abstract}
We report high resolution imaging of the nucleus of the Seyfert 1
galaxy NGC 4151 obtained with a 50 ks {\em Chandra} HRC observation.
The HRC image resolves the emission on spatial scales of $0.5\arcsec$,
$\sim$30 pc, showing an extended X-ray morphology overall consistent
with the narrow line region (NLR) seen in optical line emission.
Removal of the bright point-like nuclear source and image
deconvolution techniques both reveal X-ray enhancements that closely
match the substructures seen in the {\em Hubble Space Telescope}
[OIII] image and prominent knots in the radio jet.  We find that most
of the NLR clouds in NGC 4151 have [OIII] to soft X-ray ratio
$\sim$10, despite the distance of the clouds from the nucleus.  This
ratio is consistent with the values observed in NLRs of some Seyfert 2
galaxies, which indicates a uniform ionization parameter even at large
radii and a density decreasing as $r^{-2}$ as expected for a nuclear
wind scenario.  The [OIII]/X-ray ratios at the location of radio knots
show an excess of X-ray emission, suggesting shock heating in addition
to photoionization.  We examine various mechanisms for the X-ray
emission and find that, in contrast to jet-related X-ray emission in
more powerful AGN, the observed jet parameters in NGC~4151 are
inconsistent with synchrotron emission, synchrotron self-Compton,
inverse Compton of CMB photons or galaxy optical light. Instead, our
results favor thermal emission from the interaction between radio
outflow and NLR gas clouds as the origin for the X-ray emission
associated with the jet.  This supports previous claims that frequent
jet-ISM interaction may explain why jets in Seyfert galaxies appear
small, slow, and thermally dominated, distinct from those kpc scale
jets in the radio galaxies.
\end{abstract}


\section{Introduction}

X-ray counterparts to the powerful radio jets which extend beyond kpc,
or even Mpc, distances from radio loud active galactic nuclei (AGNs)
are well-studied (e.g., M87, Cen A, 3C273; see Harris \& Krawczynski
2006 for a review). However, their weaker analogs, the smaller jets
found on the scales of the narrow-line region (NLR) in many radio
quiet Seyfert galaxies (Nagar et al. 1999; Terashima \& Wilson 2003;
Ulvestad 2003 and references therein) are less well-studied in the
X-rays.  The limiting reasons are the angular resolution achievable in
X-rays, even with the {\em Chandra X-ray Observatory}, and the complex
circumnuclear environment often including X-ray emission contributed
from starburst and the ionized gas in the NLR (e.g., Wilson et
al. 1992; Young et al. 2001; Wilson \& Yang 2002; Wang et al. 2009)

Although challenging, studying these jets and the emission line gas
with high resolution imaging provides a valuable probe of the
interstellar medium (ISM) fueling the central engine and the
interaction between the AGN and the host galaxy (e.g., Bianchi et
al. 2006).  In particular, such high resolution data allows the
investigation of the importance of AGN jets in the energetics and
kinematics of the NLR in addition to the direct ultraviolet (UV)
emission from the nucleus.

An ideal object for such a study is NGC 4151 ($D\sim 13.3$ Mpc;
Mundell et al. 1999). It is often considered as the nearest archetypal
Seyfert 1 galaxy (see Ulrich 2000 for a review) and the nucleus
contains a linear radio jet $\sim 3.\arcsec 5$ (230 pc; Wilson \&
Ulvestad 1982; Carral et al. 1990; Pedlar et al. 1993; Mundell et
al. 1995).  The biconical NLR and the extended NLR (ENLR) are
elongated up to $\sim 10\arcsec$ along the northeast and southwest of
the nucleus and not aligned with the radio jet (Mundell et al. 2003).
The ionized gas appears clumpy in high resolution $HST$ images (e.g.,
Boksenberg et al.  1995; Winge et al. 1997; Kaiser et al. 2000).
Previous {\em Chandra} ACIS images show extended X-ray emission that
is well correlated with the optical forbidden line emission at
$r>1.\arcsec 5$ (e.g., Ogle et al. 2000; Yang et al. 2001), but cannot
investigate the association between X-ray emission and the radio jet
due to pile-up and resolution.

In this paper we present the first {\em Chandra} High Resolution
Camera (HRC) observation of the NGC 4151 nucleus.  The smaller pixel
size of HRC microchannel plate ($0.13\arcsec$ pixel$^{-1}$; {\em
  Chandra} Proposers' Observatory
Guide\footnote{\url{http://cxc.harvard.edu/proposer/POG/}}) allows
good sampling of the {\em Chandra} High Resolution Mirror Assembly
(HRMA; van Speybroeck et al. 1997; Weisskopf et al. 2002) point spread
function (PSF; FWHM$\sim0.4\arcsec$), which is instead undersampled by
the ACIS detector because of the larger physical size of CCD pixel
($0.49\arcsec$ pixel$^{-1}$).  Lack of pile-up, the superior spatial
resolution of the HRC data allows us to examine the X-ray morphology
of the nuclear region, and identify enhancements in the X-ray image
with features seen in other wavebands.

\section{Observations and Data Reduction}\

NGC 4151 was observed on 2008 March 2 starting at 10:19:48 (UT) with
the {\em Chandra} HRC-I for 50.18 ks.  The nominal pointing was
($\alpha=12^h 10^m31.^s8$, $\delta=39^{\circ}24^{\prime}33^{\prime
  \prime}$), which places the optical nucleus of the galaxy
($\alpha=12^h 10^m32.^s6$, $\delta=39^{\circ}24^{\prime}21^{\prime
  \prime}$, Clements 1981) on-axis. The total region covered was
$30^{\prime}\times 30^{\prime}$.  The HRC data were
reprocessed\footnote{\url{http://cxc.harvard.edu/ciao/threads/createL2/}}
with CIAO tool {\tt
  hrc\_process\_events}\footnote{\url{http://cxc.harvard.edu/ciao4.1/ahelp/hrc\_process\_events.html}}
using the {\em Chandra} Interactive Analysis of Observations software
package (CIAO) version 4.1 and {\em Chandra} Calibration Database
(CALDB) version 4.1.2, to generate new level 2 file that has the
latest calibration applied and the amplifier ringing effect removed.
The total exposure time was 49.67 ks after filtering of good time
intervals.

To improve the accuracy of astrometry, X-ray source detection was
performed on the HRC image using the {\tt wavdetect} algorithm
(Freeman et al. 2002) and the positions of X-ray point sources were
compared to the coordinates from the USNO-B1.0 Catalog (Monet et
al. 2003), yielding excellent absolute astrometric accuracy of
$0.2\arcsec$ (1$\sigma$).

\section{Image Analysis and Results}

\subsection{X-ray Morphology}\label{morph.sec}

Figure~\ref{fig1}$a$ presents the HRC-I image of the NGC 4151 nuclear
region, showing the central $8^{\prime\prime}\times 8^{\prime\prime}$
region.  The X-ray emission in the nuclear region is resolved into
distinct components in the HRC image, namely a bright point-like,
unresolved nucleus and resolved extended regions towards northeast
(NE) along position angle (P.A.) $\sim$48$\,^{\circ}$ and southwest
(SW) along P.A. $\sim$233$\,^{\circ}$.  The curved X-ray emission
$3\arcsec$ SW of the nucleus shows distinct segments in the HRC, which
closely follows the [OIII]$\lambda$5007 emission (e.g., $HST$/WFPC2
F502N image, Kaiser et al. 2000).  Although the ACIS and HETG zeroth
order images show similar elongation and hints of structure, some
features seen in the HRC image were not discernible due to the larger
ACIS pixel size (c.f. Figure 1 in Yang et al. 2001).  Results on new
deep ACIS imaging and detailed spectral study focusing on X-ray
emission associated with the ENLR will be presented in a separate
paper (Wang et al. 2009, in preparation).

\subsection{Preliminary Extent Analysis}\label{prep.sec}

\begin{figure}
\epsscale{1.0}
\plotone{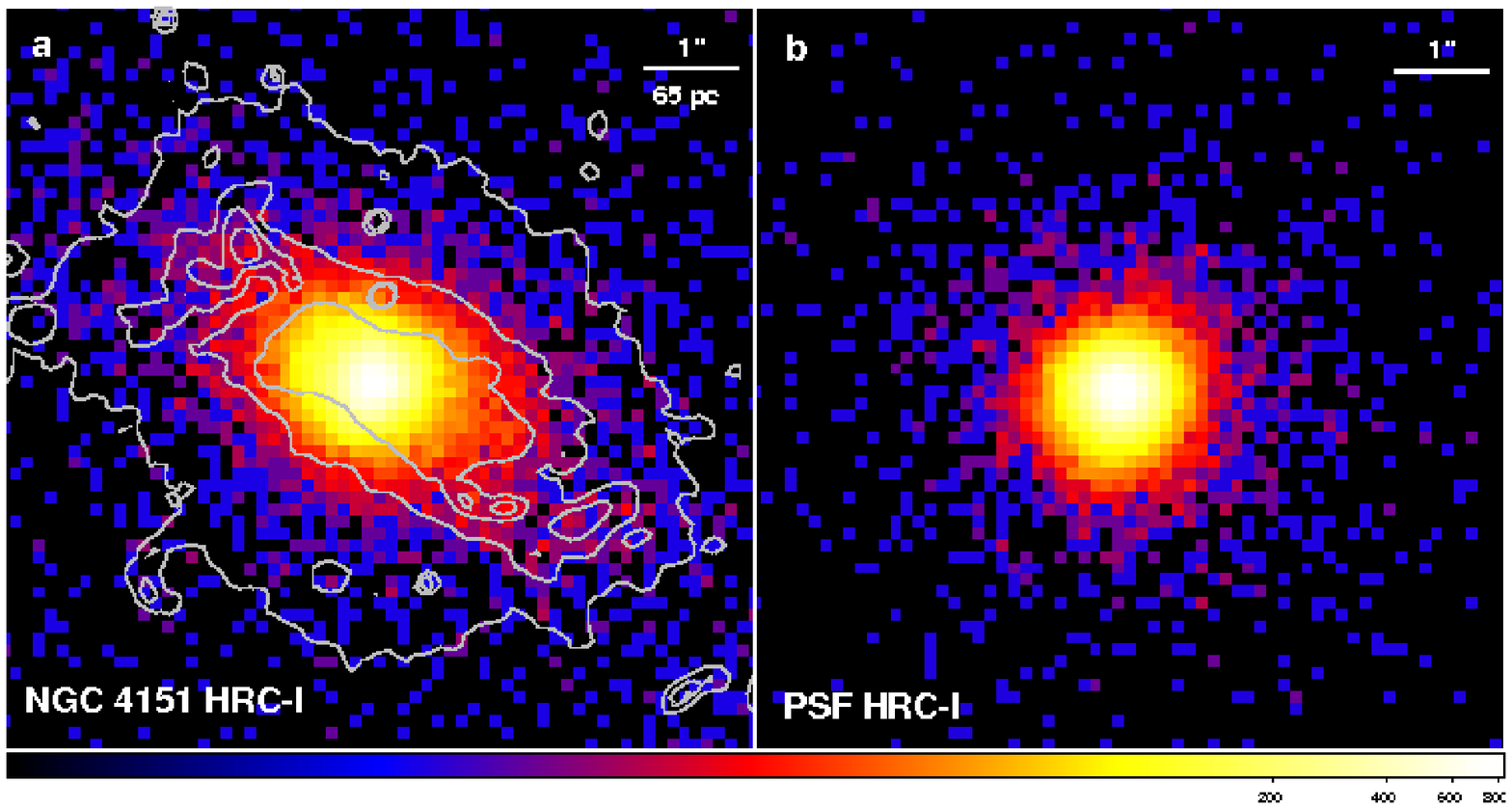}
\caption{(a) $\sim 8\arcsec\times 8\arcsec$ HRC image of the
  circumnuclear region of NGC 4151.  Contours of [OIII]5007 line
  emission are overlaid. (b) The simulated HRC PSF image. Comparing
  (a) and (b), the nuclear region of NGC 4151 clearly shows extended
  emission besides the central bright point-like source.
\label{fig1}}
\end{figure}

In order to look for low brightness emission around the bright
nucleus, we performed PSF subtraction at the nucleus position.  The
{\em Chandra} PSF was simulated with the {\em Chandra} Ray Tracer
(ChaRT\footnote{\url{http://cxc.harvard.edu/chart/}}) using a
monochromatic energy at 1 keV, sufficient for HRC data (see ChaRT
thread noted above).  The rays were then projected onto the HRC
detector with CIAO tool {\tt psf\_project\_ray} adopting a
$0.2\arcsec$ Gaussian blurring, which gives a PSF that has a sharper
radial profile matching the inner $0.5\arcsec$ data well and will be
used to perform the PSF subtraction later.  Figure~\ref{fig1}$a$ and
Figure~\ref{fig1}$b$ compare the source image and the PSF image.

\begin{figure}
\epsscale{1.0}
\plotone{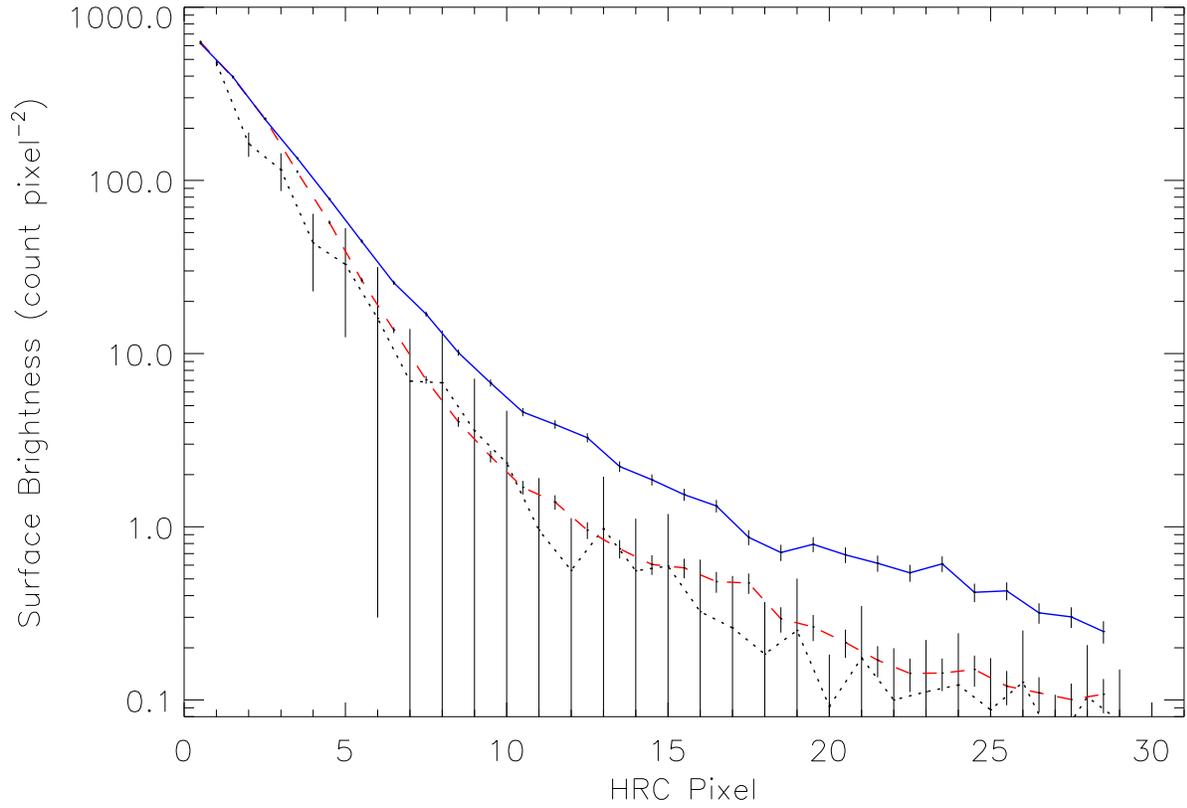}
\caption{Surface brightness profiles for all data (blue solid line),
  data within two sectors at P.A.$\sim$290$^{\circ}$--20$^{\circ}$ and
  P.A.$\sim$110$^{\circ}$--190$^{\circ}$ (black dotted line), and the simulated
  PSF (red dashed line).
\label{fig2}}
\end{figure}

In Figure~\ref{fig2}, we illustrate the presence of extended emission
along NE-SW direction by comparing surface brightness profiles.  The
nucleus radial profile deviates above the simulated PSF profile at
$r\ge 3$ pixels ($0.4\arcsec$), indicating presence of extended
emission.  In contrast, the radial profile extracted from two sectors
perpendicular to the extended emission (between
P.A.$\sim$290$^{\circ}$ and 20$^{\circ}$, and between
P.A.$\sim$110$^{\circ}$ and 200$^{\circ}$), which accurately follows
the simulated PSF profile.

From the HRC count rate, we estimate the absorption corrected 0.5--10
keV flux $F_{X,nuc.}=1.1\times 10^{-10}$ erg s$^{-1}$ cm$^{-2}$
($L_{X,nuc.}=2.3\times 10^{42}$ erg s$^{-1}$) and
$F_{X,ext.}=4.3\times 10^{-13}$ erg s$^{-1}$ cm$^{-2}$
($L_{X,ext.}=9\times 10^{39}$ erg s$^{-1}$) with PIMMS, assuming a
power law spectrum ($\Gamma=1.65$, $N_H=3\times 10^{22}$ cm$^{-2}$;
Schurch \& Warwick 2002) and a thermal bremsstrahlung spectrum
($kT=0.57$ keV, $N_H=2\times 10^{20}$ cm$^{-2}$; Yang et al. 2001) for
the nuclear point source and the extended emission (within a
$4\arcsec$ radius of the nucleus), respectively.  Note that although
these flux values agree with previous measurements in the literature
(e.g., Weaver et al. 1994, Yang et al. 2001), they rely on the assumed
spectral models and should be treated as estimates.  For example,
varying the $kT$ between 0.3 keV to 1 keV results in a $\pm 40\%$
deviation from the current flux.  Adopting the Raymond-Smith thermal
plasma model for the same $kT=0.57$ keV will increase the flux to
$4.9\times 10^{-13}$ erg s$^{-1}$ cm$^{-2}$ due to presence of strong
emission lines.

\subsection{Image with PSF Subtraction}\label{psf.sec}

\begin{figure}
\epsscale{1.0}
\plotone{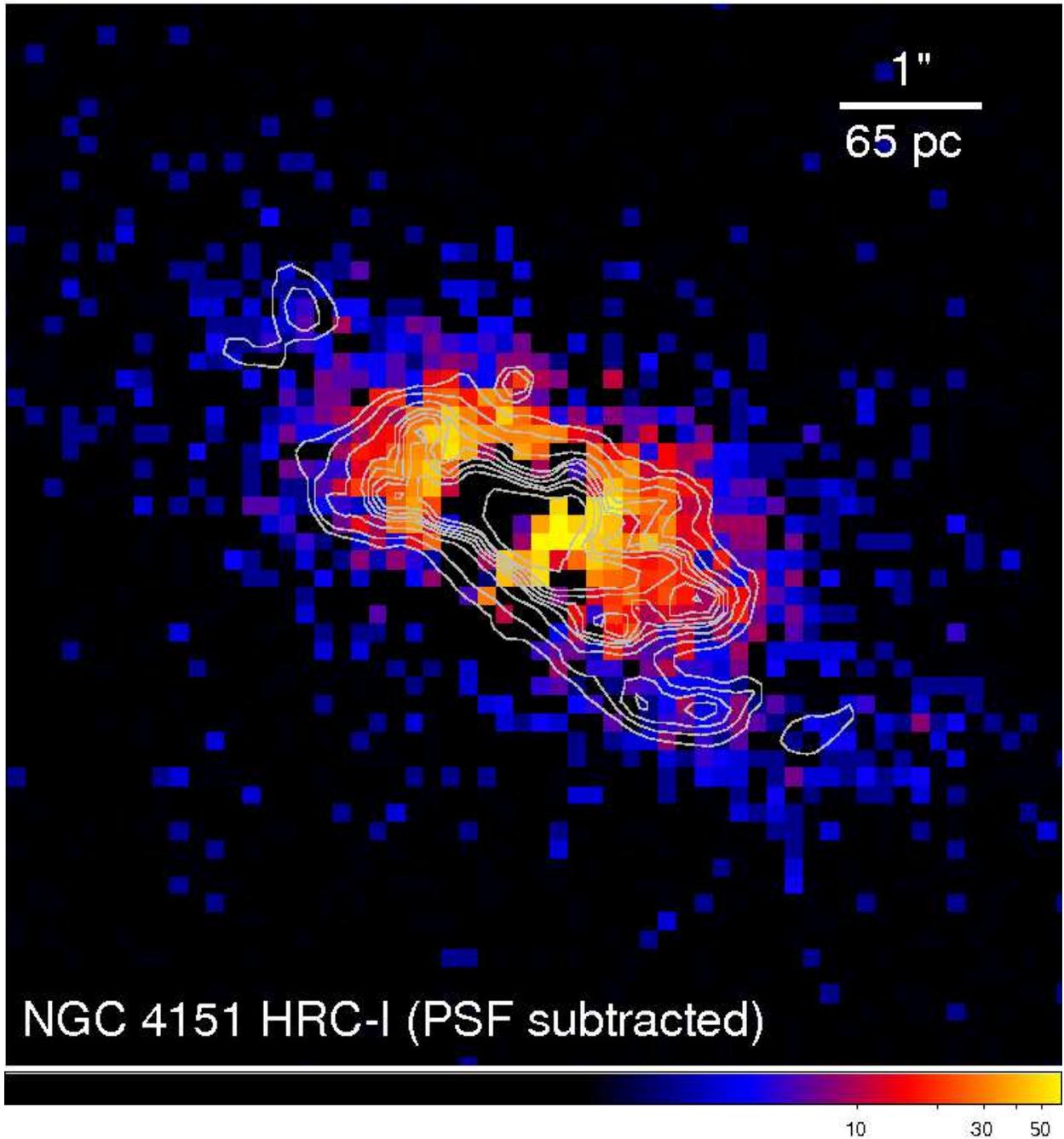}
\caption{HRC image of the nuclear region after removal of the central
  bright point-like source. Contours of $HST$/FOC F502N [OIII] line
  emission (Winge et al. 1997) are overlaid.
\label{fig3}}
\end{figure}

The peak of the PSF image is centered at the observed brightness peak
of the point-like source at detector position ($DETX$, $DETY$)=(16320,
16289), and its peak intensity is renormalized to match the point-like
source.  The resulting PSF-subtracted HRC image in the central $\sim
8\arcsec\times 8\arcsec$ region is shown in Figure~\ref{fig3} and will
be compared with the images restored with deconvolution algorithms in
\S~\ref{em.sec}.  To check how misalignments between source image and
PSF image may affect the subtraction, we offset the PSF image $\pm 1$
pixel around the observed brightness peak of the point-like source at
detector position and redid the subtraction.  There are significant
asymmetries (point source residuals) in all these offset
PSF-subtracted images with over-subtraction towards the shifted
direction and bright residual in the opposite direction, indicating
that the subtraction is off-center and our initial choice is
justified.

\subsection{Image with Deconvolution}\label{decon.sec}

To cross check the features recovered in the PSF subtracted image, we
performed image restoration with deconvolution techniques including
the widely-used Richardson-Lucy method (Richardson 1972; Lucy 1974)
and the expectation through Markov Chain Monte Carlo (EMC2; Esch et
al. 2004) method.  The Richardson-Lucy method is not well-suited for
low statistics photon-counting image, but introduces ``speckled''
appearance for extended objects (White 1994).  The EMC2 algorithm is
described in details in Esch et al. (2004) and Karovska et al. (2005)
and has well-defined convergence criteria, reliable counts and noise
estimate. It is designed to work with low count Poisson data and can
capture point sources and sharp features in the image as well as
larger scale extended features (Esch et al. 2004; Karovska et
al. 2005).  The effectiveness of this method was demonstrated with
images of astronomical objects, including interacting galaxies NGC
6240 (Esch et al. 2004), and symbiotic binary systems Mira AB
(Karovska et al. 2005) and CH Cyg (Karovska et al. 2007).

\begin{figure}
\epsscale{1.0}
\plotone{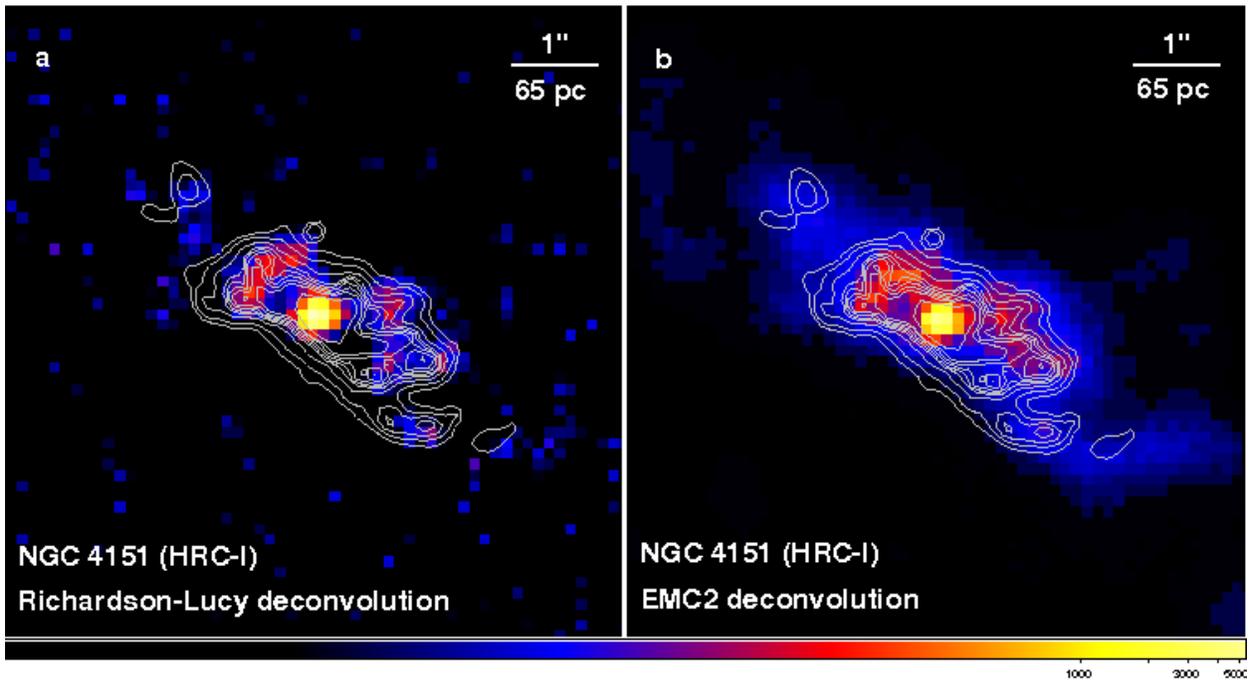}
\caption{HRC image of the nuclear region after (a) Richardson-Lucy
  deconvolution and (b) EMC2 deconvolution. Contours of $HST$/FOC F502N
  [OIII] line emission (Winge et al. 1997) are overlaid.
\label{fig4}}
\end{figure}

The restored images from both techniques are presented in
Figure~\ref{fig4}.  The Richardson-Lucy method (100-200 iterations)
gives more point-like features, while the EMC2 method (500 iterations)
shows the point-like features plus fainter extended emission.  The two
images show effectively identical X-ray enhancements along the NE-SW
direction, also share a great similarity of morphology with the
PSF-subtracted image.

\section{Discussion}\label{em.sec}

The PSF-subtracted image and deconvolved images show similar X-ray
structures, clearly indicating some relation to the NLR gas, the radio
jet, and the interaction of the jet with the ISM.  Overall, there is a
good correlation between enhancements in [OIII] and X-ray emission
(Figure~\ref{fig3} and~\ref{fig4}), possibly because both originate
from the same photoionized gas.  In the following sections, we
investigate how the X-ray emission is associated with the NLR clouds
and with the bright knots in the radio jet.

\subsection{Constraints on the X-ray Emission from the [OIII] Clouds}\label{oiii.sec}

Bianchi et al. (2006) surveyed the NLRs of 8 nearby Seyfert 2 galaxies
with $HST$ and {\em Chandra}, and found kpc-scale soft X-ray emission
coincident with the extent and morphology of the [OIII] emission.
They suggested that the same gas photoionized by the AGN continuum can
simultaneously produce the X-ray and [OIII] emission with the observed
ratios.  Note that these ratios were the average values over $\sim$kpc
regions as the ACIS images did not allow comparisons of the X-ray
emission with the small clumps seen in $HST$ images.

\begin{figure}
\epsscale{1.0}
\plotone{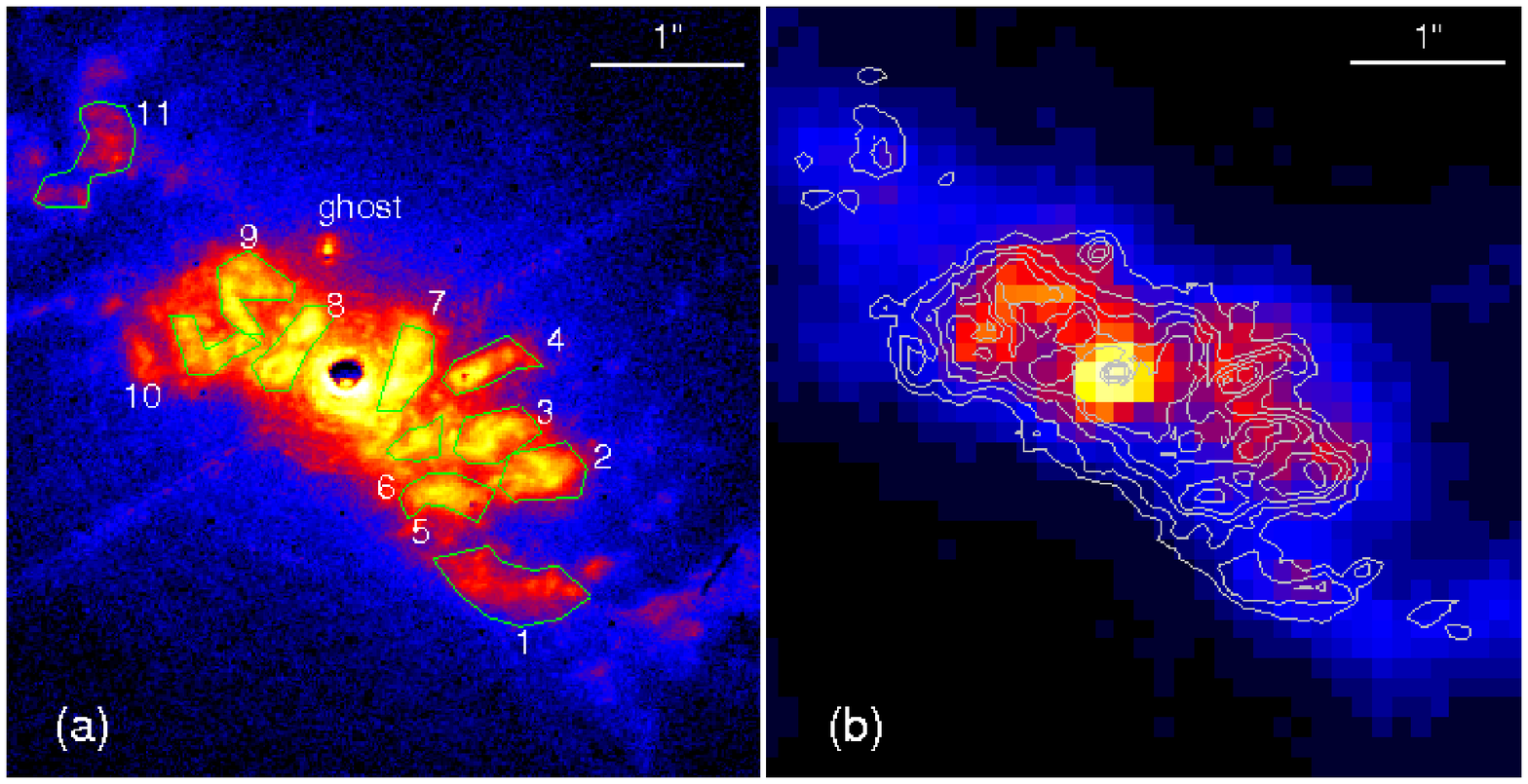}
\plotone{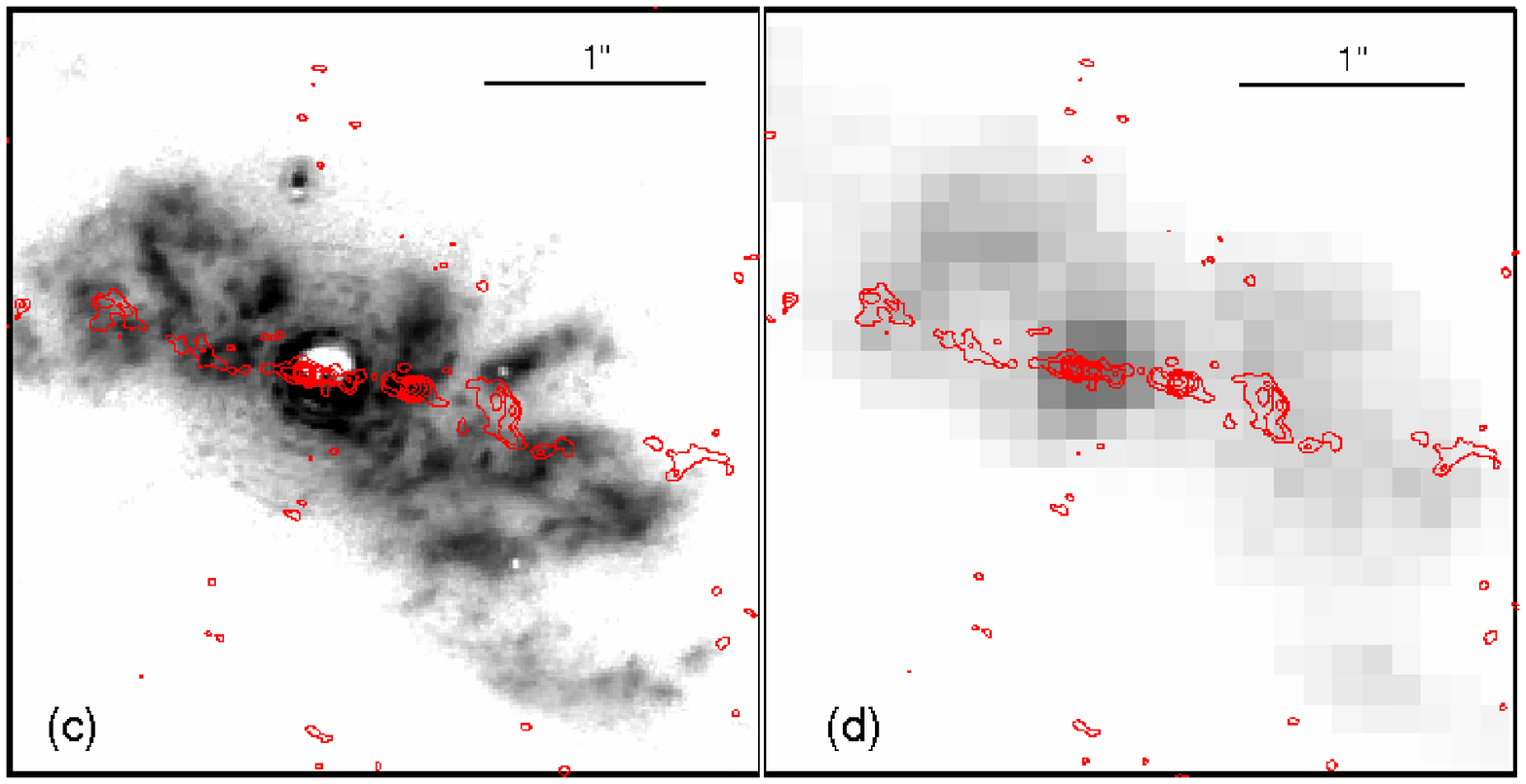}
\caption{(a) $HST$/FOC [OIII] image of the nuclear region (Winge et
  al. 1997).  The clouds listed in Table~\ref{flux} are labeled.  (b)
  Restored HRC image using EMC2 deconvolution with contours of [OIII]
  line emission from (a) overlaid. (c) and (d) shows the VLBA+VLA
  $\lambda$21 cm continuum contours (Mundell et al. 2003) overlaid on
  the $HST$/FOC [OIII] image and the HRC image with EMC2
  deconvolution, respectively.
\label{fig5}}
\end{figure}

For NGC 4151, the unprecedented high spatial resolution HRC image
enables us to compare the substructures of the X-ray emission and
those of the NLR clouds.  Figure~\ref{fig5} compares the details of a
$5\arcsec\times 5\arcsec$ $HST$ Faint Object Camera (FOC) f/96
[OIII]$\lambda 5007$ image of the nuclear region (Winge et al. 1997)
with the restored HRC image using EMC2 deconvolution.  Note the
striking correspondence of the optical [OIII] substructures to the
X-ray morphology, especially the faint cloud to the NE and a curved
extension to the SW.  The main cloud features are labeled.  Using the
calibrated FOC image, we measured the [OIII] fluxes for the clouds
following the FOC Data
Handbook\footnote{\url{http://www.stsci.edu/hst/foc/documents/foc\_handbook.html}},
and listed them in Table~\ref{flux}.  Figure~\ref{fig5} also shows the
VLBA radio image contoured on the FOC image and the HRC image with
EMC2 deconvolution, outlining the plasma flow of the spine of the jet
and the X-ray emission with respect to the jet.  We will discuss the
radio-X-ray correspondence in \S~\ref{radio.sec} and the overall
radio, [OIII], and X-ray comparison in \S~\ref{all3.sec}.

\begin{figure}
\epsscale{1.0}
\plotone{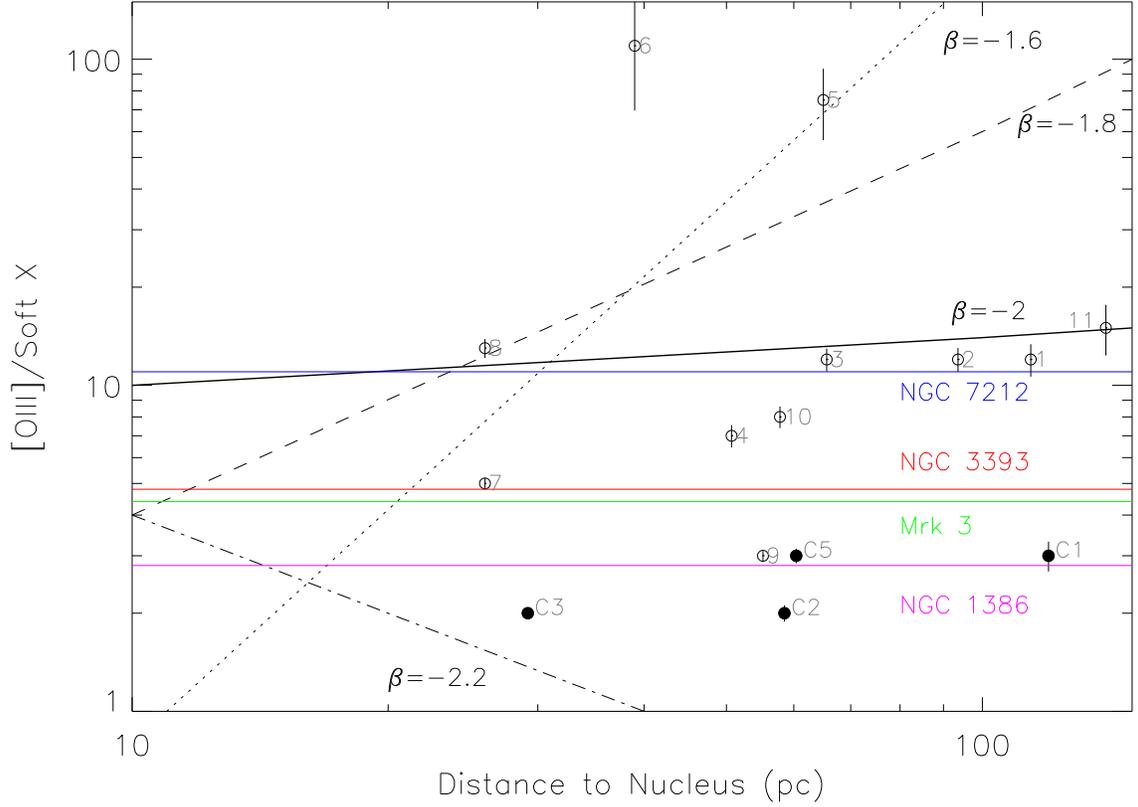}
\caption{The [OIII] to soft X-ray ratio as a function of the cloud's
  distance to the nucleus.  The open circles are the clouds as labeled
  in Figure~\ref{fig5} and Table~\ref{flux}, and the filled circles
  indicate measurements at the locations of four radio knots
  (excluding C4, the nucleus).  The blue, red, green, and magenta
  lines indicate the [OIII]/X-ray ratios for NGC 7212, NGC 3393, Mrk
  3, and NGC 1386 (Bianchi et al. 2006), respectively.  The dotted,
  dashed, solid, and dot-dashed lines are the CLOUDY model predicted
  values from Bianchi et al. (2006) for different radial density
  profiles: $n_e\propto r^{\beta}$ where $\beta=-1.6$, $-1.8$, $-2$,
  and $-2.2$, respectively.
\label{fig6}}
\end{figure}

To compare with results in Bianchi et al. (2006), the 0.5--2 keV X-ray
fluxes were also derived using counts extracted from the same regions,
with the deconvolved HRC image and PIMMS.  To check the levels of
ionization in different clouds, in Figure~\ref{fig6} we show the
[OIII] to soft X-ray ratio for the distinct cloud features
(Table~\ref{flux}) at various radii to the nucleus ($\sim$25 pc--150
pc).  

We note that two clouds (\#5 and \#6) show much higher [OIII]/X-ray
ratio ($\sim$100) than the typical value, implying lower ionization at
these locations.  Both clouds lie along the outermost edge of the SW
cone (Figure~\ref{fig5}; see also Figure~\ref{fig9}).  Comparing to
other clouds at the same radii (e.g., \#3), their lower ionization could
be explained by either a lower incident ionizing flux because of more
screening at these locations from absorbers covering the nuclear
source (see Kraemer et al. 2008), or higher density in these clouds as
they are swept up by the outflow.  It has been suggested that the NLR
could consist of different components with various degree of
ionization (e.g., Kinkhabwala et al. 2002).

In addition, the [OIII]/X-ray ratios at the 4 jet knot locations (C1,
C2, C3, C5; see \S~\ref{radio.sec} and Figure~\ref{fig7}) were
measured, and found to be uniformly low, $\sim$2.  Only one cloud
(\#9) has a similar low ratio of 3.  This implies higher X-ray
emission compared to other clouds under photoionization.  The enhanced
X-ray emission is likely associated with the outflowing radio plasma,
as many radio jets have X-ray counterparts originating from
non-thermal and thermal processes (Harris \& Krawczynski 2006).  We
explore the origin of the X-ray emission associated with these knots
in the next section.

To explain the X-ray and [OIII] ratio in a single photoionized
medium, Bianchi et al. (2006) generated a photoionization model with
CLOUDY (version 96.01, last described by Ferland et al. 1998).  We
also plot the model predicted curves in Figure~\ref{fig6} for
different radial density profiles, where the electron density was
assumed to have a power-law radial dependence $n_e\propto r^{\beta}$
($\beta=0$ is constant density, and $\beta=-2$ represents a freely
expanding wind).

We find that the 8 remaining NLR clouds in NGC 4151 have
[OIII]/X-ray(0.5--2 keV) ratio close to 10, despite the distance of
the clouds from the nucleus.  This ratio is consistent with the range
of $\sim$3-11 in the Seyfert galaxies observed by Bianchi et
al. (2006), although it is at the higher end.  The fairly constant
[OIII]/X-ray ratios indicate an almost uniform ionization parameter
even at large radii, requiring a density dependence close to $r^{-2}$,
as expected for a wind from the nucleus (e.g., Krolik \& Kriss 1995;
Elvis 2000).  This agrees with the conclusion in Bianchi et al. (2006)
and the results found for some well-studied NLRs (e.g., Kraemer \&
Crenshaw 2000; Collins et al. 2005; Kraemer et al. 2008).

\subsection{Constraints on the X-ray Emission from the Radio Jet Knots}\label{radio.sec}

\begin{figure}
\epsscale{1.0}
\plotone{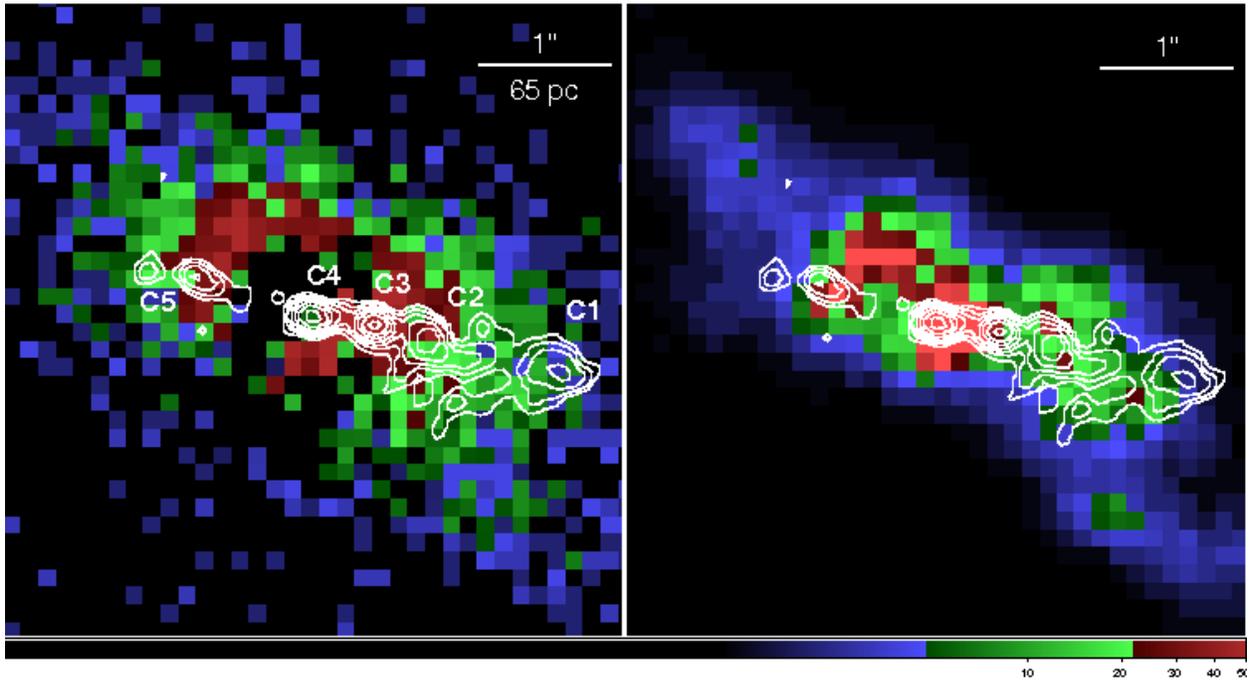}
\caption{Zoom-in image of the inner $4\arcsec$ of the NGC 4151 nuclear
  region: (a) PSF subtracted image, and (b) restored image using EMC2
  deconvolution. The overlays are contours from MERLIN 1.4 GHz radio
  map (Mundell et al. 1995; $HPD=0.15\arcsec$) and the main jet
  components C1-C5 are labeled.
\label{fig7}}
\end{figure}

Figure~\ref{fig7} shows the inner $\sim 4\arcsec\times 4\arcsec$
region of the PSF subtracted image and the restored image using EMC2
deconvolution, overlaid with contours of the radio jet (MERLIN 1.4 GHz
map, Mundell et al. 1995).  There are five main radio components in
the radio jet (C1-C5) from a number of studies (Carral et al. 1990;
Pedlar et al. 1993; Mundell et al. 1995).  C4 is known as the position
of the nucleus and was used to align the X-ray peak. In jet components
C3 and C5, the radio knot and X-ray enhancement appear to originate in
the same volume at current resolution of the X-ray image.  On the
other hand, C2 has little X-ray emission but is straddled by two X-ray
blobs and coincides with jet deflections in ``Z''-like shape (see
Figure~\ref{fig7} and also Figure~\ref{fig5}).

To understand the association of X-ray emission with the jet knots, we
consider the following emission mechanisms in the general framework of
X-ray emission processes in radio jets (Harris \& Krawczynski 2006).
The spectral index of a power law $\alpha$ is defined by flux density,
$S_{\nu}\propto \nu^{\alpha}$ following the radio convention.  To
evaluate the X-ray flux densities of the knots, we extracted HRC
counts from the EMC2 deconvolved image using regions defined by radio
contour ($3\sigma$ at 1 mJy), which have comparable resolution ($\sim
0.15\arcsec$, Mundell et al. 1995).  We also attempted to extract HRC
counts from the PSF subtracted image using regions defined by
resampling radio contours to match the lower HRC resolution, which
yielded similar counts (excluding the nucleus, C4).  The HRC has poor
energy resolution and the data are not amenable to standard spectral
fitting, therefore the X-ray flux is estimated over 0.1-10 keV range
with PIMMS (see Table~\ref{simple} footnote for details) assuming a
power law index $-\alpha=1.0$, a typical value in low radio power jets
(Harris \& Krawczynski 2006) and Galactic-only absorption
($N_H=2\times 10^{20}$ cm$^{-2}$; Yang et al. 2001).  Radio flux
densities of the knots were taken from Pedlar et al. (1993).

\begin{figure}
\epsscale{1.}
\plotone{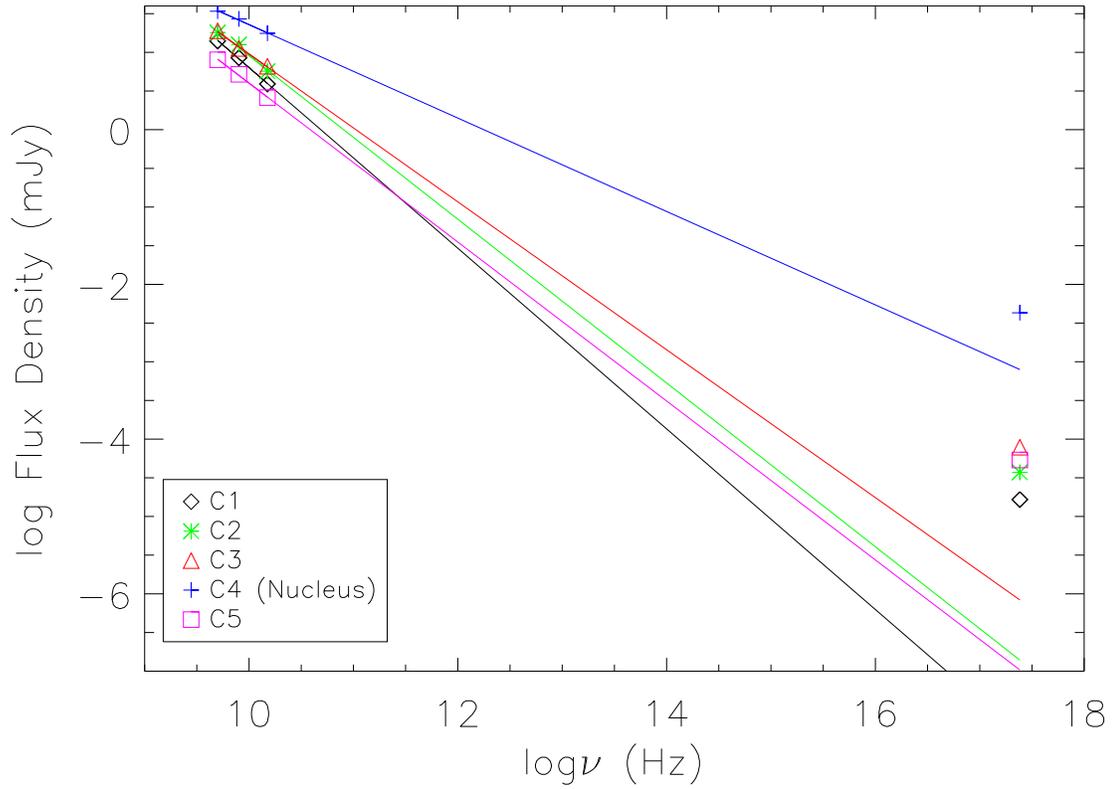}
\caption{Spectra of radio knots.  The X-ray emission is taken from
  Table~1 assuming a power law index $-\alpha=1$.  Symbols: C1 (black
  diamond), C2 (green asterisk), C3 (red triangle), C4 (blue plus), C5
  (magenta square).  The error bar is negligible at this scale.  The
  lines with correponding colors are linear fit of the slopes for the
  radio points.
\label{fig8}}
\end{figure}

\begin{figure}[hbt]
\epsscale{1.0}
\plotone{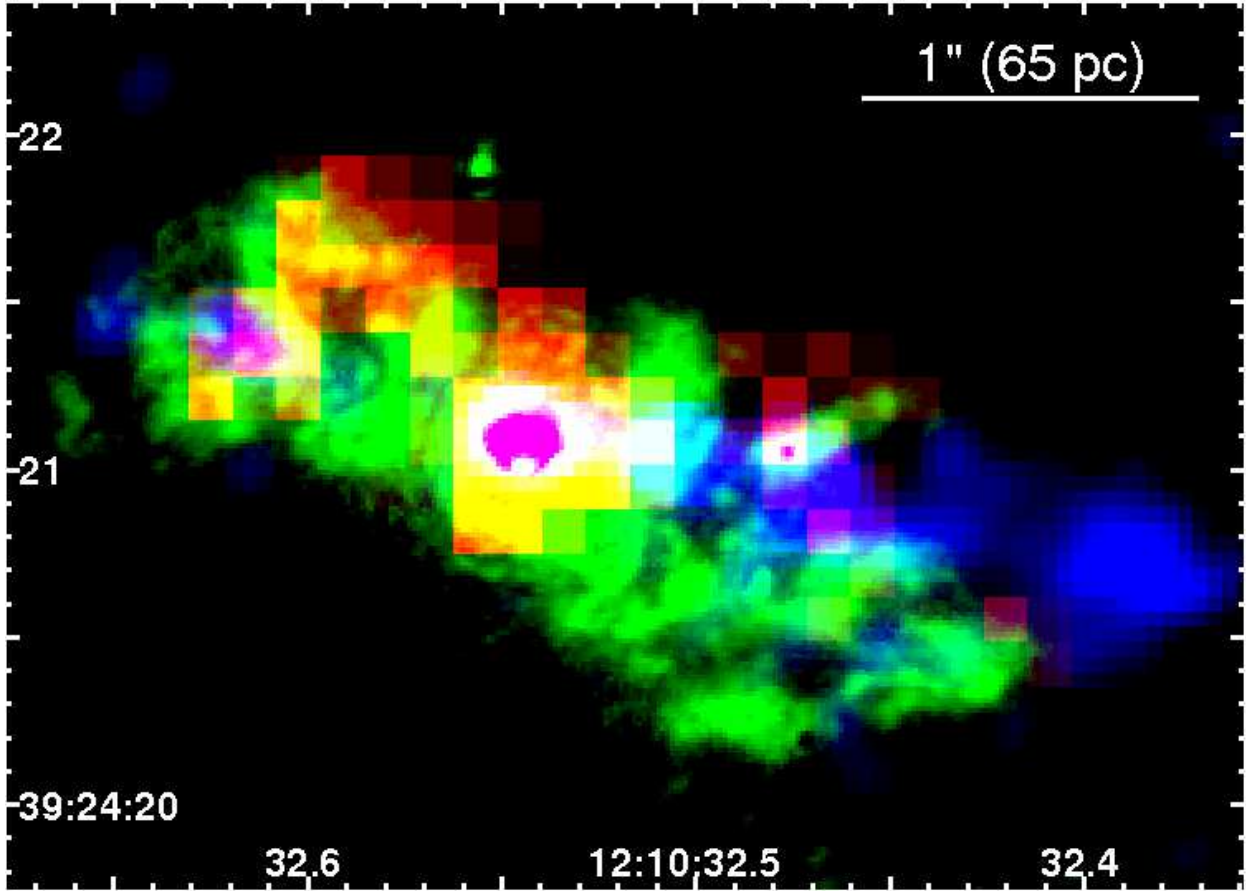}
\caption{Composite image of the NGC 4151 nuclear region. Blue is the
  MERLIN radio map (Mundell et al. 1995), green represents optical
  [OIII] line emission image (Winge et al. 1997) and red is the HRC
  image after PSF deconvolution.  The X-ray counts lower than 10 are
  blanked to show the strongest features.
\label{fig9}}
\end{figure}

{\it Synchrotron Emission} -- In many radio jets, circumstantial
evidence exists for the synchrotron process generating the X-rays in
the knots.  If this is the dominant process in the NGC 4151 jet, the
X-ray intensity would be consistent with a single power law
extrapolation or a broken power law concaving downward.
Table~\ref{simple} gives the emission parameters of the radio
components.  Following the minimum energy argument (or equipartition)
generally adopted for synchrotron sources (e.g., Govoni \& Feretti
2004), we listed in Table~\ref{density} the typical magnetic fields in
the knots, most of which have $B\approx 1$ mG, assuming a
proton-to-electron ratio $K=100$ (Pedlar et al. 1993).
Figure~\ref{fig8} shows spectra of the radio knots.  The observed
X-ray intensities lie orders of magnitude above the extension of the
radio synchrontron spectra.  This cannot be attributed to the
uncertainties in the radio or X-ray flux density measurements, which
indicate that a simple synchrotron model is insufficient for the X-ray
emission.

{\it Inverse Compton Emission} -- Low frequency photons are scattered
by relativistic electrons to higher frequency through the inverse
compton (IC) process.  One common emission process in radio jets is the
synchrotron self-compton (SSC) emission.  The photon energy density
from the synchrotron spectrum in each knot can be calculated, using
$u_{sync}=3L_{sync} R/4cV$ (Wilson et al. 2000), and assuming
uniformly emitting spheres to derive volumes, where $L_{sync}$ is the
radio luminosity, $R$ is the sphere radius, $c$ is speed of light, and
$V$ is the volume.  Another common IC process in radio jets is the IC
scattering of the cosmic microwave background (CMB), which has the
photon energy density $u_{CMB}=4\times 10^{-13}$ ergs cm$^{-3}$.

Both $u_{sync}$ and $u_{CMB}$ are much lower than photon energy
density of the combined AGN and star light in the NGC 4151 nuclear
region, therefore considering the latter as the dominant seed photons
is more appropriate.  To simplify the estimate, we approximate the
photon field as blackbody radiation peaking at $T=4000$ K, with a
energy density $u_{ph}=2$ ergs cm$^{-3}$.  Following Blumenthal \&
Gould (1970), we derive an estimate of the magnetic field from the
ratio between the X-ray and radio fluxes.  For all the cases, the
required $B$ values for IC mechanism to explain the X-ray emission are
$\sim 3$ orders of magnitude larger than the equipartition magnetic
field $B\sim 1$ mG, which is unlikely.

To ease the requirement on the magnetic field, beaming model with
relativistic bulk jet velocity must be invoked to boost IC emission.
However, there is strong evidence against a highly relativistic bulk
velocity of NGC 4151 jet.  First, the angle between the jet and our line
of sight is $\sim 40^{\circ}$ (Pedlar et al. 1993).  The knots that
have similar distance to the nucleus (C2 and C5) also have comparable
X-ray/radio intensities. The fact that we see a two-sided, non-boosted
radio jet suggests the bulk velocity is not highly relativistic.
Secondly, Ulvestad et al. (2005) measured the speeds of the jet component
with VLBI and found 0.05$c$ and 0.028$c$ at 0.16 and 6.8 pc from the
nucleus, respectivelly, confirming the non-relativistic jet motions.
None of the forms of IC emission can account for the observed X-ray
fluxes.

{\it Thermal Bremsstrahlung Emission} -- The X-ray emission from the
radio features may originate from hot gas rather than from non-thermal
mechanism, although with the current resolution we cannot distinguish if
the hot gas is located within the radio emitting volume or around the
jet.  Considering the morphology, jet-cloud interaction seems to be
present.  We adopted a $kT\sim 0.6$ keV for the thermal model as a low
$kT$ is typical of the X-ray emission from the NLR ionized gas and
measured from the X-ray spectral fitting (Yang et al. 2001).  In
Table~\ref{density} we calculated the emission measure, electron
number density, thermal pressure for each knot.  It is often argued
that the absence of Faraday rotation and depolarization places a limit
on the required electron densities (e.g., de Young 2002). In NGC 4151
we derive $n_e<10^{11}$ cm$^{-3}$ from optical polarization
measurement of Kruszewski (1971), assuming an equipartition field
$B\sim 10^{-3}$G.  The $n_e$ required for thermal emission is orders
of magnitude smaller than this limit. A thermal origin is thus highly
plausible.

\subsection{X-ray, Radio, and [OIII] Comparison}\label{all3.sec}

Figure~\ref{fig9} compares altogether X-ray emission (red) with the
radio jet (blue; MERLIN 1.4 GHz map, Mundell et al. 1995) and optical
NLR emission (green; $HST$/FOC F502N [OIII]$\lambda 5007$ image, Winge
et al. 1997) in projection.  The radio component C4 contains the AGN
(see higher resolution VLBA studies by Ulvestad et al. 2000; Mundell
et al. 2003).  Assuming the peak of the optical nuclear emission
originates from the AGN, we aligned the X-ray, optical and radio
nuclei.

There are X-ray enhancements associated with the bright radio knots in
the jet as well as the NLR clouds (see also Figure~\ref{fig7}).  The
overall morphology in the three bands is consistent with the scenario
that clumpy material lies in the path of the jet and is shock-heated
to X-ray emitting temperature from the impact with the outflowing
radio plasma from the nucleus.  C1 has largely diffuse morphology in
the radio and weak X-ray emission; it is mostly in an [OIII]-emission
cloud free region.  Around knots C2 and C5, as noted in Mundell et
al.(2003), a number of [OIII] clouds are closely associated with the
radio knots and appear to bound the radio knots (see
Figure~\ref{fig5}c).  The morphologically disturbed radio jet may have
cleared a path through the NLR (Mundell et al. 2003).

Some evidence has been reported supporting this jet-cloud interaction
scenario.  Kinematic studies mapping the full velocity field of the
NLR clouds (Winge et al. 1999; Kaiser et al. 2000) found that the jet
may be influential in producing the high velocity dispersions for the
clouds in the inner $4\arcsec$, although not directly responsible for
the acceleration of the gas (Crenshaw et al. 2000; Das et al. 2005).
Mundell et al. (2003) examined the apparent radio correspondence with
these clouds, suggesting that sites of radio jet deflection are
aligned in projection with high velocity dispersion clouds.

Storchi-Bergmann et al. (2009) mapped near infrared emission-line
intensities and ratios in the NLR of NGC 4151, which probe the effects
of shocks produced by the jet on the NLR gas.  We note that there are
enhancements of the [Fe II] emission at the locations of radio knots
C2 and C5 in these IR maps (e.g., the [Fe II]/Pa$\beta$), consistent
with being the spots of jet-cloud interaction.  As a minimal
requirement of the shock scenario, following Kraft et al. (2009) the
pressure of the knots must be less than the ram pressure of the jet,
which translates to $p_{knot}< 2P_{jet}/v_j A$ ($p_{knot}$ is the
pressure of a knot, $P_{jet}$ is the jet power, $A$ is the
cross-section area, and $v_j$ is the jet velocity).  Using an
estimated jet power $P_{jet}\sim 1.6\times 10^{43}$ ergs s$^{-1}$
(Allen et al. 2006), $v_j=0.028c$ (Ulvestad et al. 2005), and $A=30$
pc, we find $p_{knot}<10^{-6}$ dyne cm$^{-2}$, which is satisfied by
our derived pressure (Table~\ref{density}).

Our estimates for the NGC 4151 jet assuming thermal origin of the
X-ray emission match well with the characteristics of the jets in
hydrodynamical simulations (e.g., Rossi et al. 2000, Saxton et
al. 2005), which are relatively heavy ($\rho> 1$ cm$^{-3}$) and slow
($v<5\times 10^4$ km s$^{-1}$).  This is also consistent with the
conclusion of Whittle et al. (2004, 2005) studies of a jet-dominated
Seyfert Mkn 78, where a thermally-dominated, slow and dense Seyfert
jet encountering dense gas clouds was identified.

Such jet-cloud interaction may explain why the jets in Seyfert
galaxies seem very different from those in radio-loud AGNs (Middelberg
et al. 2007): they are not able to propagate freely as do the
well-collimated, galactic scale jets in radio galaxies.  Besides NGC
4151, there is strong evidence for interactions of radio jets with the
ISM on the scales of NLRs in Seyfert galaxies (e.g., NGC 1068, Wilson
\& Ulvestad 1982; IC 5063, Oosterloo et al. 2000; NGC 2110, Evans et
al. 2006; III Zw 2, Brunthaler et al. 2005; NGC 3079, Middelberg et
al. 2007), which would be worth {\em Chandra} follow-up imaging to
locate the X-ray emission.

It is also worth noting that the NE part of the X-ray emission (e.g.,
cloud \#9) appears brighter than the SW part.  According to the
modeled geometry of the bicone of ionized gas and host galaxy in Das
et al. (2005), the SW side is closer to us and our line of sight is
outside of the bicone.  One plausible explanation for the enhanced
X-ray emission in NE is that, the NE bicone intersects with the NE
galactic disk and the X-ray emitting medium there may have higher
density.  On the other hand, although SW part of the bicone also
intersects the disk, our line of sight to the intersection goes
through the cone and may be subject to higher absorption (see Figure
10 in Evans et al. 1993).

As a cautionary note, physical association between the features seen
in different bands along the line of sight is not warranted because of
the projection effect.  At the knot position, at least part of the
X-ray emission could be contributed from the NLR clouds directly
ionized by the AGN (e.g., Bianchi et al. 2006).  But as we showed in
\S~\ref{oiii.sec}, C1, C2, and C5 are not associated with any bright
NLR clouds.  Instead, thermal X-ray emission is expected in the
jet-ISM interaction scenario described above, which is well supported
by the low [OIII]/soft-X ratios and the enhancement of [FeII],
together with the multiwavelength morphologies.

\section{Conclusions}

The high resolution imaging of the NGC 4151 nucleus obtained with {\em
  Chandra} HRC shows X-ray morphology that is both overall consistent
with the NLR seen in optical line emission, with substructures closely
matching the [OIII] clouds, and with knots in the radio jet, implying
X-ray emission associated with both the photoionized gas and the jet
components. 

We find that most of the NLR clouds in NGC 4151 have [OIII] to soft
X-ray ratio $\sim$10, at or a factor of $\sim$10 in distance of the
clouds from the nucleus.  The radially constant ratio indicates a
uniform ionization parameters even at large radii and a density
dependence $\propto r^{-2}$ as expected for a nuclear wind.  

The calculations of required jet parameters from observed X-ray and
radio properties do not favor synchrotron emission, SSC emission, or
IC of CMB photons and the local galaxy light.  Thermal emission from
interaction between radio outflow and the NLR clouds is the most
favorable explanation. 

Future high spatial resolution X-ray observatories, such as {\em
  Generation-X} with an angular resolution of $0.1\arcsec$ (Brissenden
2009), will be able to unambiguously resolve the X-ray emission from
the jets and the NLR with high spectral resolution and so gain
important new knowledge of the outflows of both thermal and
non-thermal plasma from Seyfert galaxies.

\acknowledgments

We thank the anonymous referee for useful comments that improved the
clarity of our paper.  This work is partially supported from NASA
grant GO8-9101X and NASA Contract NAS8-39073 (CXC).  We are grateful
to Dan Harris and Aneta Siemiginowska for their stimulating discussion
on radio jets.  J. W. thanks E. Galle and M. Juda (CXC) for technical
assistance in HRC data reduction.

{\it Facilities:} \facility{CXO (HRC, ACIS)}

\clearpage

\clearpage

\clearpage

\clearpage

\clearpage

\clearpage

\clearpage

\clearpage

\clearpage

\clearpage

\begin{deluxetable}{cccccc}
\tabletypesize{\scriptsize}
\tablecaption{Measured X-ray and [OIII] Fluxes\label{flux}}
\tablewidth{0pt}
\tablehead{
\colhead{\begin{tabular}{c}
Cloud\\
label\\
\end{tabular}} &
\colhead{\begin{tabular}{c}
Distance\\
to Nuc. ($\arcsec$)\\
\end{tabular}} &
\colhead{\begin{tabular}{c}
Distance\\
to Nuc. (pc)\\
\end{tabular}} &
\colhead{\begin{tabular}{c}
[OIII] flux\\
($10^{-13}$erg s$^{-1}$ cm$^{-2}$)\\
\end{tabular}} &
\colhead{\begin{tabular}{c}
0.5-2 keV flux\\
($10^{-14}$erg s$^{-1}$ cm$^{-2}$)\\
\end{tabular}} &
\colhead{\begin{tabular}{c}
[OIII]/soft X\\
\end{tabular}}
}
\startdata
1 & 1.76 & 114 & 1.1 & 0.91$\pm0.10$ & 12\\
2 & 1.44 & 93.6 & 1.9 & 1.61$\pm0.13$ & 12\\
3 & 1.01 & 65.6 &2.2 & 1.84$\pm0.14$ & 12\\
4 & 0.78 & 50.7 &1.3 & 1.91$\pm0.14$ & 7\\  
5 & 1.0 & 65.0 &1.5 & 0.23$\pm0.05$ & 75\\
6 & 0.6 & 39. &1.1 & 0.12$\pm0.04$ & 110\\
7 & 0.4 & 26. &3.4 & 7.38$\pm0.29$ & 5\\
8 & 0.4 & 26. &3.3 & 2.60$\pm0.17$ & 13\\
9 & 0.85 & 55.2 &2.4 & 9.59$\pm0.33$ & 3\\
10 & 0.89 & 57.8 &1.6 & 2.00$\pm0.15$ & 8\\
11 & 2.15 & 139.7 &0.6 & 0.45$\pm0.07$ & 15\\
\hline
C1 & 1.84 & 119.6 &0.3 & 1.14$\pm0.11$ & 3\\
C2 & 0.9 & 58.5 &0.8 & 3.58$\pm0.21$ & 2\\
C3 & 0.45 & 29.2 &3.3 & 13.9$\pm0.40$ & 2\\
C5 & 0.93 & 60.4 & 1.4 & 4.49$\pm0.22$ & 3\\
\enddata
\tablecomments{The knot containing nucleus C4 is not included here.}
\end{deluxetable}

\clearpage
\begin{deluxetable}{cccccccc}
\rotate
\tabletypesize{\scriptsize}
\tablecaption{Parameters of the Jet Features\label{simple}}
\tablewidth{0pt}
\tablehead{
\colhead{\begin{tabular}{c}
Knot\\
\end{tabular}} &
\colhead{\begin{tabular}{c}
Counts\\
\end{tabular}} &
\colhead{\begin{tabular}{c}
$F_{0.3-10{\rm{keV}}}$\\
($10^{-12}$ erg s$^{-1}$ cm$^{-2}$)\\
\end{tabular}} &
\colhead{\begin{tabular}{c}
S[1 keV]\\
(mJy)\\
\end{tabular}} &
\colhead{\begin{tabular}{c}
S[5 GHz]\\
(mJy)\\
\end{tabular}} &
\colhead{\begin{tabular}{c}
S[8.4 GHz]\\
(mJy)\\
\end{tabular}} &
\colhead{\begin{tabular}{c}
S[15 GHz]\\
(mJy)\\
\end{tabular}} &
\colhead{\begin{tabular}{c}
$-\alpha$ (5/8GHz)\\
\end{tabular}}
}
\startdata
C1 & 143 & 0.14 & 0.04 & 14& 8.5& 3.9 & $-$1.0\\
C2 & 316 & 0.32 & 0.09 & 18& 12.6& 5.7& $-$0.9\\
C3 & 1224 & 0.64 & 0.19  & 19& 11.2& 6.6& $-$1.0\\
C4 & 23342 & 100. & 10.4 & 34& 27& 17.6 & $-$0.4\\
C5 & 390 & 0.39 & 0.13 & 8& 5.2& 2.6 & $-$0.9
\enddata

\tablecomments{A fixed Galactic column $N_H=2\times 10^{20}$ cm$^{-2}$ is
  applied to all. $N_H=3\times 10^{22}$ cm$^{-2}$ is assumed for an extra
  absorption column towards the nucleus (C4; Yang et al. 2001). The
  X-ray flux is derived assuming a spectral index $-\alpha=1.0$,
  representative of X-ray jets in low power radio galaxies (Harris \&
  Krawcznski 2006).}
\end{deluxetable}

\clearpage

\begin{deluxetable}{cccccccc}
\rotate
\tabletypesize{\scriptsize}
\tablecaption{Radiation Fields of the Jet Features\label{density}}
\tablewidth{0pt}
\tablehead{
\colhead{\begin{tabular}{c}
Knot\\
\end{tabular}} &
\colhead{\begin{tabular}{c}
Radio size ($\arcsec$)\\
\end{tabular}} &
\colhead{\begin{tabular}{c}
$U_{sync}$ (erg cm$^{-3}$)\\
\end{tabular}} &
\colhead{\begin{tabular}{c}
$B_{eq}$ (G)\\
\end{tabular}} &
\colhead{\begin{tabular}{c}
$B_{IC}$ (G)\\
\end{tabular}} &
\colhead{\begin{tabular}{c}
E.M. ($\times 10^{63}$ cm$^{-3}$)\\
\end{tabular}} &
\colhead{\begin{tabular}{c}
$n_e$ (cm$^{-3}$)\\
\end{tabular}} &
\colhead{\begin{tabular}{c}
$P$ (dyne cm$^{-2}$)\\
\end{tabular}}
}
\startdata
C1 & 0.5 & $8.7\times 10^{-15}$  & $1.3\times 10^{-3}$ & 3.1 & 0.8 & 14.6 & $2.4\times 10^{-8}$\\
C2 & 0.5 & $1.1\times 10^{-14}$  & $1.3\times 10^{-3}$ & 1.6 & 2.1 & 22.1 & $3.4\times 10^{-8}$\\
C3 & 0.2 & $7.3\times 10^{-14}$  & $4.8\times 10^{-3}$ & 1.2 & 7.8 & 194.9 & $3.1\times 10^{-7}$\\
C5 & 0.3 & $1.4\times 10^{-14}$  & $1.5\times 10^{-3}$ & 0.6 & 1.1 & 53.1 & $8.4\times 10^{-8}$
\enddata
\tablecomments{$U_{sync}$ is the energy density of synchrontron emission. $B_{eq}$ is the equipartition magnetic field strength. $B_{IC}$ is the magnetic field required to produce the observed X-ray emission through IC process. E.M. is the thermal emission measure of the hot gas. $n_e$ is the electron number density. $P$ is the thermal pressure of the hot gas.  The knot containing nucleus, C4 is not included here.}
\end{deluxetable}

\end{document}